\documentclass[11pt]{article}
\usepackage{moriond,epsfig}

\def\be{\begin{equation}}
\def\ee{\end{equation}}
\def\bea{\begin{eqnarray}}
\def\eea{\end{eqnarray}}

\def\etal{{\it et al.\ }}

\begin{document}
\vspace*{4cm}
\title{A REVIEW OF THE HIGH-REDSHIFT SUPERNOVA SEARCHES}

\author{ A.G. KIM }

\address{PCC, Coll{\`e}ge de France, 11, Place M. Berthelot,\\
75231 Paris, France}

\maketitle\abstracts{
Observations show that Type Ia Supernovae (SNe Ia)
form a homogeneous class of objects.  They share similar
spectroscopic evolution, light-curve shapes, and peak absolute
magnitudes.
The slight departures
from homogeneity that are observed
can be used to
produce a ``calibrated candle'' with
corrected magnitudes with even smaller dispersion.
The existence of this intrinsically bright distance
indicator has inspired two coordinated high-redshift supernova
searches: the Supernova Cosmology Project and the High-z
Supernova Search Team.
To date $\sim 100$ SNe Ia have been discovered by the two groups.
The preliminary analysis of the first of these objects demonstrate how well
SNe Ia can be used to measure
the mass density of the universe $\Omega_M$ and
the normalized cosmological constant $\Omega_\Lambda \equiv \Lambda/3H_0^2$.
}

\section{Introduction}
For the past several years, two independent groups have been discovering
and following high-redshift supernovae ($z>0.3$) using telescopes from all over
the world and beyond.  The lofty and imposing goal of these searches?
To determine the ultimate fate of the universe!  But before I tell you what
the answer is (so far), I should explain what makes supernovae so special
and give you an idea of what's been observed to date.  Then comes the
answer, along with a discussion
of some of the systematic errors involved
and how we can address them.  I conclude by presenting the scientific
course we plan to take in the near future.

\section{Type Ia Supernovae as Distance Indicators}
One of the big reasons that supernovae are exciting for cosmologists
is because of the remarkable homogeneity
of the type Ia's (SN Ia)
\footnote{The supernova
classification system is empirical and is based on the
spectrum, SNe Ia exhibit no hydrogen (hence the ``I'') and have strong silicon
P-Cygni features (hence the ``a'') during their photospheric phase.}.
This homogeneity is seen in their evolving spectra (Filippenko\cite{fi:araa}),
their light-curve shapes (Leibundgut\cite{le:thesis}),
and their peak absolute magnitudes
which have a dispersion
of $\sigma \approx 0.3 - 0.5$ mag
depending on the sample (Branch and Tammann\cite{br:araa}).
SNe Ia are also whoppingly bright, at peak they can emit
as much light as their host galaxy.  This combination of
brightness and homogeneity
means that they can be used to measure distances
out to very high redshifts.

The standard model for the SNe Ia progenitor system has
a white dwarf in a binary system
accreting matter from its companion until
it reaches the Chandrasekhar mass ($\sim 1.4M_\odot$)
triggering a thermonuclear runaway which we observe as a supernova.
This idea neatly explains the homogeneity and the lack of hydrogen in the
spectra.
Detailed theoretical work has provided strong support for this model
(Nomoto \etal \cite{nomoto:1994}) although there are a few outstanding
questions that have to be resolved, for example
the nature of the binary companion and
the hydrodynamics of flame propagation.

Strong evidence for intrinsic {\it in}homogeneity first came with the
light curves of the Cal{\'a}n-Tololo supernova search.  It was shown that
the light-curve shape was correlated with the supernova peak brightness
in such a way that the slow decliners tend to be brighter than the rapid
decliners (Hamuy \etal \cite{hamuy:1996a}; Riess, Press, and
Kirshner\cite{ri:lcs}).  The slow decliners were also
bluer in the optical passbands (Riess, Press, and Kirshner\cite{ri:lcs2})
and had a
much stronger UV flux (Branch, Nugent, and Fisher\cite{br:aig}).
The line ratios of particular spectral features at maximum light
also vary with light-curve shape, an effect that has been modeled
as being due to differences in the supernova's photospheric
temperature (Nugent \etal \cite{nu:1995}).

We can take advantage of this inhomogeneity by using relations between
the absolute magnitude and these other independent observables to produce
a ``calibrated candle'' with an even tighter absolute
magnitude dispersion than
before.  Such corrections using light-curve shapes yield corrected
magnitude dispersions of $\sigma \simeq 0.18$ mag.

Note that
${\cal M} \equiv M-5\log H_0=m-5 \log(cz)$
is the ``absolute magnitude'' accurately
measured for supernovae in the Hubble flow.  It is not sensitive to
the uncertainty in the Hubble constant and so is used
in lieu of the true absolute magnitude.

\section{The Searches}
There are currently
two independent teams that are running coordinated search
and follow-up observations of high-redshift supernovae.  They are
the Supernova Cosmology Project (SCP) (Perlmutter \etal \cite{pe:nature})
which was launched in 1989,
and the High-Z Supernova
Search Team (HIZ) (Garnavich \etal \cite{ga:1998}) which found their first
supernova in 1995.
Both teams use similar techniques (described in detail
in Perlmutter \etal \cite{pe:aig})
and resources
as follows.

A special strategy is needed to find these rare and random events.
Several days after new moon, a series of wide-field images are taken
on a 4-m class telescope, with each field containing over a thousand galaxies
that can potentially host a supernova.  Several weeks later the same fields
are re-observed and scanned for new point sources.  After applying cuts
to reject asteroids, AGN, quasars, cosmic rays, and other sources of
background, we are left with
supernova candidates.
Having run the search right before new moon we have optimal observing
conditions for
pre-scheduled spectroscopy to identify the candidates,
and photometry to build their multi-band light curves.
The three week gap in the search
is well matched for the $\sim 20$ day (rest frame)
rise time for SNe Ia, meaning that most all the supernovae
will be discovered before or at maximum light.
With the allocated search time (typically a pair of two nights)
both groups have been yielding
$\sim 12$ supernovae per run.

Most of the current searching is performed on the 30'x30' field of the
BTC at the CTIO 
Blanco Telescope.
With its 10-m diameter collecting area, the Keck Telescope is the
spectroscopic workhorse for both groups, allowing us to efficiently observe
and confirm
a large number of faint candidates.
Photometric follow-up is performed at a host of 2 to 4-meter telescopes
all around the world.

\begin{figure}[p]
\begin{center}
\psfig{figure=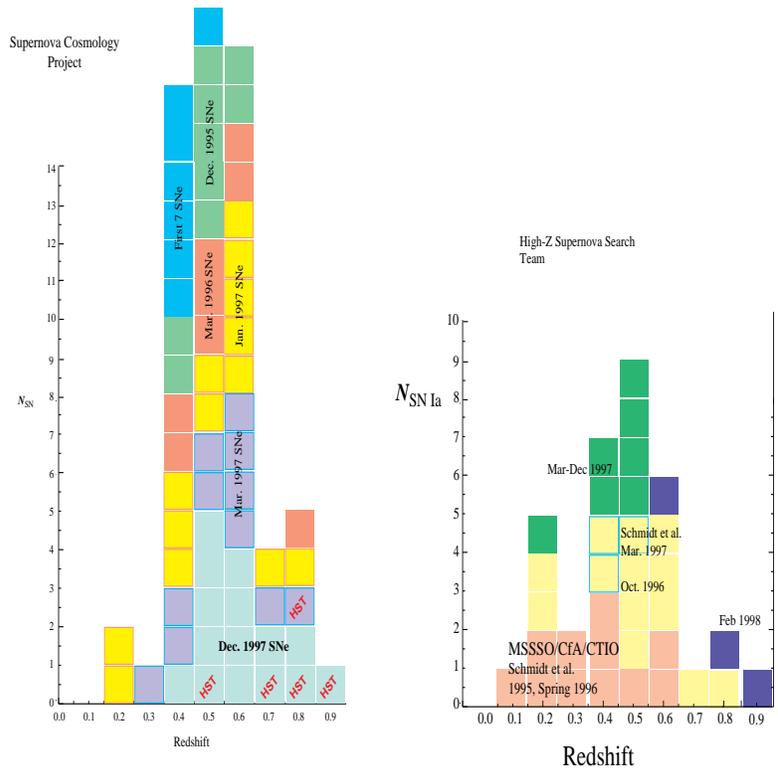,width=4in,height=4in}
\end{center}
\caption{The redshift distributions of the SNe Ia found by the two searches.
Note that four of the first seven
SCP supernovae shown do not have spectroscopic
confirmation.}
\label{hist}
\end{figure}

In total, there have been $\sim 100$ SNe Ia with spectral
confirmation discovered at $z>0.3$ by
the two groups.
Histograms describing their redshift distributions are given in Figure
\ref{hist}.
The mean redshifts of the discovered supernovae have been
steadily increasing with each successive search run,
as we have specifically tailored the filters
and exposure times to search at progressively larger distances.

Even though we have been chugging along doing practically the same thing
for several years, there are a couple of new developments that I find
exciting.
We now have found 4 supernova with
$z>0.8$ and we expect
that number to grow quickly.  By pushing our detectors
to the limit, we could search even deeper than we are now.
(The furthest supernova to date with a confirmed SN spectrum is SN1998I
at $z=0.89$, discovered by the HIZ group.)
Both groups have scheduled HST time for photometric follow-up, giving us
precise photometry and host morphology identification.  Furthermore,
at higher redshifts a supernova's rest-frame optical light
reaches us in the infra-red.  The HST NICMOS camera (while it lasts)
gives us a superior view
in this wavelength regime than would be possible from the ground.
The first results using the HST data have already been published
(Perlmutter \etal \cite{pe:nature}; Garnavich \etal
\cite{ga:1998}).

\section{Measurement of $\Omega_M$ and $\Omega_\Lambda$}
By using SNe Ia as distance indicators,
we can measure the cosmological parameters.
The standard Friedmann-Lema{\^\i}tre
cosmology gives a magnitude-redshift relation
as a function of $\Omega_M$ and
the normalized cosmological constant $\Omega_\Lambda \equiv \Lambda/3H_0^2$:
\begin{equation}
m_R(z) =  {\cal M}_B+5\log({\cal D}_L(z;\Omega_M,\Omega_\Lambda))+K_{BR},
\label{answer}
\end{equation}
where
$K_{BR}$ is the K correction relating $B$ magnitudes of nearby
SNe with
$R$ magnitudes of distant objects (Kim, Goobar, and Perlmutter\cite{kim:kcorr})
and the ``absolute magnitude'',
${\cal M}_B$, is determined using local supernovae
in the Hubble flow.
(Recall
that it is ${\cal M}_B$ that depends on the light curve shape.)
Here we use ${\cal D}_L$, the ``Hubble-constant-free'' part of the
luminosity distance, $d_L$:
\begin{equation}
{\cal D}_L(z;\Omega_M,\Omega_\Lambda) \equiv d_LH_0=
\frac{c(1 + z)}{\sqrt{|\kappa| }} \; \; \; {\cal S}\! \left (
\sqrt{|{\kappa}| } \int_0^{z} \left [(1+z^\prime)^2(1+\Omega_M z^\prime)-
z^\prime (2+z^\prime ) \Omega_\Lambda \right]^{-\frac{1}{2}} dz^\prime
\right ),
\label{R}
\end{equation}
where for $\Omega_M + \Omega_\Lambda > 1$,
${\cal S}(x)$ is defined as $\sin(x)$ and $\kappa
= 1 - \Omega_M - \Omega_\Lambda $;
for $\Omega_M + \Omega_\Lambda < 1$, ${\cal S}(x) =
{\rm sinh}(x)$ and $\kappa$ as above; and for $\Omega_M + \Omega_\Lambda = 1$,
${\cal S}(x) = x$ and
$\kappa =1$, where $c$ is the speed of light in units of
${\rm km~s^{-1}}$.

What these equations show is that the difference between
the absolute and observed magnitudes of a supernova at a given redshift
corresponds with a strip (a line if we don't include uncertainties)
in the $\Omega_M$ -- $\Omega_\Lambda$ plane.  Furthermore, the shape and
orientation of the band
are different at different redshifts, meaning that
supernova measurements from a wide range of redshifts have confidence
regions whose
intersection gives a closed area
in the $\Omega_M$ -- $\Omega_\Lambda$ plane
(Goobar and Perlmutter\cite{goo:lambda}).
This allows us to
make a simultaneous measurement of $\Omega_M$ and $\Omega_\Lambda$
with the added bonus the answer does not depend on the Hubble constant.

Both groups have published results from the first handful of supernovae
found (Perlmutter \etal \cite{sn1992bi}; Perlmutter \etal \cite{pe:1997};
Perlmutter \etal \cite{pe:nature}; Garnavich \etal \cite{ga:1998}).
I will present here the preliminary results from the analysis of the first
$\sim 40$ supernovae from SCP.  Results from the HIZ collaboration are
presented in Leibundgut's paper in these proceedings and a new paper based
on their first $\sim 15$ events is in the works.

\begin{figure}[p]
\begin{center}
\psfig{figure=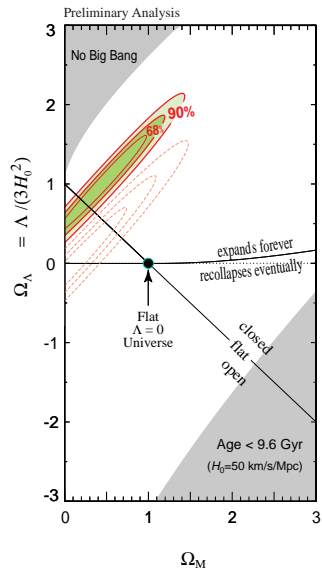,height=3in}
\end{center}
\caption{Preliminary confidence regions from the first $\sim 40$ SCP
supernovae.  The confidence regions are strictly statistical,
the lower contours show the results if there is
a 0.2 magnitude systematic difference between local and distant supernovae.}
\label{conf}
\end{figure}

Figure~\ref{conf} shows the preliminary confidence regions in the
$\Omega_M$ -- $\Omega_\Lambda$ plane for the first $\sim 40$ SCP supernovae.
The length of the region shows that at the moment
we cannot simultaneously constrain $\Omega_M$ and $\Omega_\Lambda$ down to an
interesting level because
we currently lack a large number of $z>0.8$ supernovae.
The skinniness of the region is due to the reduced statistical
error from the large number of supernovae
observed at $z \sim 0.5$.  It allows us to make statistically
significant measurements
of the cosmological parameters if we assume a flat
($\Omega_M+\Omega_\Lambda=1$)
universe, $\Omega_M=0.025 \pm 0.06 \pm 0.3$, or a $\Lambda=0$ universe,
$\Omega_M=-0.4 \pm 0.1 \pm 0.5$, where the first error is statistical
and the second is an estimate of systematic error.
Of profound interest is the fact that our supernovae
strongly disfavor the flat $\Lambda=0$ universe predicted by the
simplest theories of inflation.
Fortunately the HIZ team is getting similar results.

\section{Systematics}
Its clear that our results are now limited by systematic errors and that
these errors need to be seriously addressed.
The confidence regions shown in Figure~\ref{conf} are calculated using
only our statistical uncertainty.  To illustrate the effect of systematic
errors, we plot a second set
of contours in Figure~\ref{conf}, which show how our confidence region would
shift if high-redshift supernovae were systematically 0.2 mag fainter
than the nearby calibrators, (while maintaining the same light-curve shapes).
Systematic errors generally have the effect of shifting our contours in
the $\Omega_M$ -- $\Omega_\Lambda$ plane, smearing the confidence of our
measurement.  Detailed descriptions of how we handle some of these systematics
are given in Perlmutter \etal \cite{pe:1997}

Malmquist bias is a fancy way of saying that in a magnitude limited
sample, we are more likely to observe intrinsically brighter objects.
For a given redshift,
this produces a shift in the mean observed magnitude as compared to the
intrinsic mean.
Such an effect in our distant supernova would cause us to
overestimate $\Omega_M$ so corrections for this effect would moves us even
further from a $\Lambda=0$ flat universe.
To measure the influence of Malmquist bias,
we have determined the detection efficiencies
and thresholds for our search and performed a fit using only the
subsample of supernovae found far from the detection limit;
no statistically significant change in $\Omega_M$--$\Omega_\Lambda$
is seen.  More disturbing is the fact
that the effect of Malmquist bias on the local calibrators
is not easily calculable.  Many of the nearby supernovae were found either
randomly or on photographic plate searches.  A correction for
Malmquist bias in the nearby sample  would move us closer to
a $\Lambda=0$ flat universe.
What is now needed is a determination of
the intrinsic population of SNe Ia from a large sample
of local supernovae found in CCD searches with known detection efficiencies.
The similar distributions currently seen
in high-z and local light-curve shapes
do indicate there is no large relative difference in the effect of
Malmquist bias in the two samples.

There is no guarantee that local and distant supernovae are exactly the
same.  One may expect systematic differences in the
progenitor system metallicity or in the
white dwarf $C/O$ ratio.  Our most powerful test for evolution
is with the comparison of distant and nearby supernova
spectra, because in spectra we should see the effects
of different initial progenitor compositions
(H{\"o}flich, Wheeler, and Thielemann\cite{hoflich:1997}).
Fortunately no such
spectral redshift evolution has been noted in the spectra
observed to date, e.g. SN1997ap at $z=0.83$
(Perlmutter \etal \cite{pe:nature}).
Important information could also come from supernovae at
$0.1<z<0.3$ because it is feasible to obtain spectral
time series which could show the first effects of redshift evolution.

In order to see the effect of extinction, we
have compared the k-corrected color distributions of the local and distant
supernovae.  The negligible difference between the two
indicates that the samples have similar $E(B-V)$
distributions.
We also directly fit $E(B-V)$ for each supernova using the multi-band data
and again find similar distributions for the two sets.
Work is currently being performed to try
to reduce the errors involved in individual supernova corrections.
The fairly large uncertainty in $B-V$ for supernova at
maximum as a function of light-curve shape translates into a large
correlated uncertainty in magnitude absorption. In addition the
extinction properties
($R_B$) at high-redshift are not well known.



There are population effects that can be seen in local
supernovae.  For example, supernovae in ellipticals on whole have
skinnier light curves and are fainter than their spiral
counterparts.  The magnitude--light-curve shape relation in principle
corrects such effects.  However,
considering all the possible
population dependencies,
we would like to select local and distant supernovae subsets
that
share as many population characteristics as possible.
The steady stream of new
nearby supernovae coming from a wide range of environments
will allow us to do this.

We currently believe that we understand the effects of most of these
systematics, at least to first order.
We await new data from
high-quality nearby supernovae
searches to confirm our findings and to give us
even stronger constraints on our errors.
Specifically,
the new data will supply us with the intrinsic supernova
luminosity function
and a large homogeneous set of light curves that start before maximum.
The data will also provide us with
large sample from which we can select subsamples with which to explore
systematic errors.

\section{Conclusion}
We can confidently say that our data disfavors
an $\Omega=1$, $\Lambda=0$ universe, considering the huge amount of
systematic error required to make the two consistent.
For the future, we need to pursue two opposite directions
in order to get a more precise value of the cosmological parameters.
By finding more supernovae at $z>0.8$, we will attempt to
reduce the length of the contour in Figure~\ref{conf} to give a better
simultaneous measurement of $\Omega_M$ and $\Omega_\Lambda$.  By finding
supernovae at $z < 0.2$ we will learn more about their intrinsic properties
and give a larger sample with which we can study and hopefully
reduce systematic effects.
SCP is now dedicating a large part of its efforts in
nearby supernova searching,
and in collaboration a group of French scientists is planning to use
the CFHT specifically for $z>0.8$ searches.
(HIZ's most distant candidate
was in fact found using the CFHT).
Members of the HIZ team are already heavily involved in nearby searches
and as a whole are pursuing higher redshifts.
With all this focused activity, reduction of the systematic and statistical
errors
should not be far off in the future.


\section*{References}
\begin{thebibliography}{99}

\bibitem{fi:araa}
A.~V. Filippenko.
\newblock {\em ARAA}, 35:309, 1997.

\bibitem{le:thesis}
B.~Leibundgut.
\newblock PhD thesis, University of Basel, 1988.

\bibitem{br:araa}
D.~Branch and G.~A. Tammann.
\newblock {\em ARAA}, 30:359, 1992.

\bibitem{nomoto:1994}
K.~Nomoto, H.~Yamaoka, T.~Shigeyama, S.~Kumagai, and T.~Tsujimoto.
\newblock In {\em Supernovae : Les Houches, session LIV )}, page 199.
  North-Holland:Amsterdam, 1994.

\bibitem{hamuy:1996a}
M.~Hamuy, M.~M. Phillips, N.~B. Suntzeff, R.~A. Schommer, J.~Maza, and
  R.~Aviles.
\newblock {\em AJ}, 112:2391, 1996.

\bibitem{ri:lcs}
A.~G. Riess, W.~H. Press, and R.~P. Kirshner.
\newblock {\em ApJ}, 438:L17, 1995.

\bibitem{ri:lcs2}
A.~G. Riess, W.~H. Press, and R.~P. Kirshner.
\newblock {\em ApJ}, 473:88, 1996.

\bibitem{br:aig}
D.~Branch, P.~Nugent, and A.~Fisher.
\newblock In {\em Thermonuclear Supernovae}, page 715. Dordrecht:Kluwer, 1997.

\bibitem{nu:1995}
P.~Nugent, M.~Phillips, E.~Baron, D.~Branch, and P.~Hauschildt.
\newblock {\em ApJ}, 455:L147, 1995.

\bibitem{pe:nature}
S.~Perlmutter \etal
\newblock {\em Nature}, 391:51, 1998.

\bibitem{ga:1998}
P.~M. Garnavich \etal
\newblock {\em ApJ}, 493:L53, 1998.

\bibitem{pe:aig}
S.~Perlmutter \etal
\newblock In {\em Thermonuclear Supernovae}, page 749. Dordrecht:Kluwer, 1997.

\bibitem{kim:kcorr}
A.~G. Kim, A.~Goobar, and S.~Perlmutter.
\newblock {\em PASP}, 108:190, 1996.

\bibitem{goo:lambda}
A.~Goobar and S.~Perlmutter.
\newblock {\em ApJ}, 450:14, 1995.

\bibitem{sn1992bi}
S.~Perlmutter \etal
\newblock {\em ApJ}, 440:L41, 1995.

\bibitem{pe:1997}
S.~Perlmutter \etal
\newblock {\em ApJ}, 483:565, 1997.

\bibitem{hoflich:1997}
P.~{H{\"o}flich}, J.~C. Wheeler, and F.~K. Thielemann.
\newblock {\em astro-ph/9709233}, 1997.

\end{thebibliography}
\end{document}